\documentclass[preprint,12pt]{elsarticle}

\usepackage{amssymb}
\usepackage{color}

\journal{}

\begin{document}

\begin{frontmatter}



\title{A generalized voter model with time-decaying memory on a multilayer network}


\author{Li-Xin Zhong$^{a,b}$}\ead{zlxxwj@163.com}
\author {Wen-Juan Xu$^c$}
\author {Rong-Da Chen $^{a,b}$}
\author{Chen-Yang Zhong$^d$}
\author {Tian Qiu$^e$}
\author {Yong-Dong Shi$^f$}
\author {Li-Liang Wang$^g$}

\address[label1]{School of Finance and Coordinated Innovation Center of Wealth Management and Quantitative Investment, Zhejiang University of Finance and Economics, Hangzhou, 310018, China}
\address[label2]{China Academy of Financial Research, Zhejiang University of Finance and Economics, Hangzhou, 310018, China}
\address[label3]{School of Law, Zhejiang University of Finance and Economics, Hangzhou, 310018, China}
\address[label4]{Yuanpei College, Peking University, Beijing, 100871, China}
\address[label5]{School of Information Engineering, Nanchang Hangkong University, Nanchang, 330063, China}
\address[label6]{Research Center of Applied Finance, Dongbei University of Finance and Economics, Dalian, 116025, China}
\address[label7]{Hangzhou Dianzi University, Hangzhou, 310018, China}

\begin{abstract}
By incorporating a multilayer network and time-decaying memory into the original voter model, the coupled effects of spatial and temporal cumulation of peer pressure on consensus are investigated. Heterogeneity in peer pressure and time-decaying mechanism are both found to be detrimental to consensus. The transition points, below which a consensus can always be reached and above which two opposed opinions are more likely to coexist, are found. A mean-field analysis indicates that the phase transitions in the present model are governed by the cumulative influence of peer pressure and the updating threshold. A functional relation between the consensus threshold and the decaying rate of the influence of peer pressure is found. As to the time to reach a consensus, it is governed by the coupling of the memory length and the decaying rate. An intermediate decaying rate may lead to much lower time to reach a consensus.

\end{abstract}


\end{frontmatter}


\section{Introduction}
\label{sec:introduction}

In the physical world, the occurrence of phase transitions between a disordered system and an ordered system has long been an important issue which has been extensively studied by scientists\cite{liu2,wang3,Johnson1,han1,li1,schneidman1}. For example, in the magnetic field, the spin-up and spin-down dynamics have been investigated by statistical physicists\cite{havlin2,zhou1,zheng1,zhou2,Zhong10,Zhong11}. The spatial-temporal conditions for the magnetic order-disorder transition have been found. As far as a coordination between two simultaneous
processes is concerned, D. Li et al. have investigated the synchronization of coupled phase oscillators\cite{li10}. The interfaces between different domains of synchronized oscillations have been found. Similar phenomena related to the order-disorder transition have also been found in social and economic fields\cite{perc1,schweitzer1,lu1,zhou3,qian1}. In a social context, competing for universal agreement is quite important in some typical situations, such as presidential election and public affairs decision-making. The conditions for reaching consensus have been deeply explored by a variety of researchers, including sociologists, economists, statistical physicists and mathematicians\cite{stauffer1,hadzibeganovic1,torok1,zhong1}.

In the study of opinion dynamics, the binary-choice voter model and the Hegselmann-Krause model represent two distinctive kinds of opinion formation processes\cite{fortunato1,fortunato2,vazquez1,vazquez2,qu1,xu1}. In the voter model, an individual's opinion has two possible states, labelled $+1$ and $-1$ respectively. Depending upon some kind of updating rule, the individuals update their states sequentially or simultaneously. In the Hegselmann-Krause model, the individuals hold diverse opinions, which are expressed as real numbers within the range of [-1,+1]. According to a bounded confidence rule, the individuals gradually modify their opinions. J. Shao et al. have studied the nonconsensus opinion model on different networks\cite{shao1}. They have found that the clustering of the individuals with the same opinion can prevent themselves from being invaded by the individuals with opposite opinion. J. Fernandez-Gracia et al. have investigated the driving forces for the occurrence of the statistical features of U.S. presidential elections\cite{juan1}. They have found that social influence, mobility, and population heterogeneity have great impacts on the fluctuation features of vote share.

The opinion dynamics provides us an insight into how a consensus can be reached within an interacting group. So far, in the study of the dynamics of opinion formation, great efforts have been made to study the effects of comparatively stable interpersonal relationship and real-time interactions between different individuals. However, in the real world, an individual may have a diverse set of interpersonal relations\cite{groeber1,blattner1,johnson3,lorenz1,li2}. He may belong to different kinds of social communities, such as relative relationship, colleague relationship and business relationship. The overlapping communities in complex networks have been discussed in ref.\cite{li10}. In addition to that, different communities usually have different impacts on one's opinion formation. What the Matthew effect demonstrates is such an inequality of advantages. Related studies on preferential attachment have been reviewed by M. Perc in ref.\cite{perc10}. In addition to the diversity in communities, another scenario in opinion dynamics is the non-real-time response. A real-time interaction may not result in an immediate change but a time-lagged change of one's behavior\cite{granell1,juher1}. The coupled effects of diverse interpersonal relations and the cumulative influence of individual interactions on opinion formation need an in-depth study.

The multi-community interactions are usually modelled as a multi-layer network\cite{gomez1,buldyrev1,diakonova1,nicosia1}. By incorporating a two-layer network into the SIR model, L. G. Alvarez Zuzek et al. have investigated the isolation effect on epidemic spreading\cite{zuzek1}. They have found that the epidemic threshold increases with the rise of the isolation period. S. Havlin et al. have studied the percolation transition on interdependent networks\cite{havlin1}. They have found that the strong coupling of subnetworks leads to discontinuous phase transitions. M. Perc et al have studied the evolution of cooperation on dynamic networks and multilayer networks\cite{perc2,wang1,wang2}. They have found that the equilibrium and non-equilibrium processes are greatly affected by the interaction structures. Z. Wang et al. have studied interdependent network reciprocity in public goods game\cite{wang20}. They have found that only when the coordination between different subnetworks is not disturbed, can cooperation be promoted. The coevolutionary dynamics of network interdependence and game strategy has been investigated by Z. Wang et al.\cite{wang21}. They have found that the coevolutionary system can reach optimal conditions for cooperation.  Recent researches related to multi-layered systems, including structures, dynamics, evolution and emergent properties have been reviewed by D. Y. Kenett et al. in ref. \cite{kenett10}.

The cumulative influence is usually modelled as memory. By incorporating memory of past exposures to contagious influence into the original SIR model, P. S. Dodds et al. have proposed a generalized epidemic model\cite{dodds1}. Three classes of epidemic dynamics have been found. C. Liu et al. have studied the memory effect on information propagation\cite{liu1}. Compared with the situation where only internal or external influence exists, the coexistence of them makes the information spread more quickly and broadly. In relation to the time-lagged effect. The decaying confidence mechanism has been introduced into the opinion formation model\cite{arescu1}. It has been found that local agreement is easier to be achieved than global consensus. H. U. Stark et al. have investigated the effect of time-dependent transition rate on opinion formation \cite{stark1,stark2}. They have found that an intermediate inertia rate may lead to much lower time to reach a consensus. The memory effects on the evolution of social ties and spreading processes have been extensively studied in references\cite{xu2,medus1,rosvall1,hou1,wang10,bao1}.

Inspired by the studies on the multi-layer network model and the epidemic model with memory\cite{buldyrev1,dodds1,stark1}, in the present work, we introduce a generalized voter model into which a multilayer network and memory of past influences are incorporated. The spatial-temporal conditions for consensus are extensively investigated. The following are our main findings.

(1)Heterogeneity in peer pressure is detrimental to consensus. On condition that the peer pressure is homogeneous. The effect of increasing multilayer is similar to the effect of increasing random connections. The more communities each individual participates in, the easier it is to reach a consensus. On condition that the peer pressure is heterogeneous. Increasing  multilayer is similar to reducing the mean  influence of peer pressure. The critical value of multilayer, below which increasing  multilayer promotes consensus while above which increasing multilayer leads to a nonconsensus state, is found.

(2)The time-decaying mechanism does harm to consensus and deepens the heterogeneous influence of peer pressure. Increasing the decaying rate of the influence of peer pressure reduces the maximal value of the cumulative influence of peer pressure, which leads to a decrease in the transition point of the updating threshold. The coupling of the heterogeneity factor $\Delta\omega$ and the decaying rate $\xi$ of the influence of peer pressure results in a smaller value of the transition point $d_{th}^c$ of the updating threshold. A functional relation between $d_{th}^c$ and $f(\Delta\omega,\xi)$ is found.

(3)The convergence time is typically affected by the heterogeneity in peer pressure and the decaying rate. For a small value of memory length, an increase in $\Delta w$ or $\xi$ leads to a continuous rise of the convergence time. For a large value of memory length, an increase in $\xi$ firstly leads to a decrease and then a sharp rise of the convergence time.A functional relation between the convergence time and the coupling of the updating probability and the synchronicity of the updates of two opposed opinions is found.

A generalized voter model with time-decaying memory on a multi-layer network is introduced  in section 2. Numerical results are presented and discussed in section 3. Theoretical analyses are given in section 4 and conclusions are drawn in section 5.

\section{The model}
\label{sec:model}

The generalized voter model is defined as follows. There is a multilayer network consisting of $L$ subnetworks. In each subnetwork, there are $N$ nodes which are randomly connected with probability $p$. Therefore, the average degree of each node in the subnetwork satisfies the equation $<k>=(N-1)p$. There exist $N$ individuals. Each one belongs to $L$ subnetworks and stays at a typical node of each subnetwork. Such linkage denotes that each individual in the present model belongs to $L$ different kinds of communities, which implies that an individual's behavior should be affected by all the individuals in these communities. The influence of $L$ communities on one's opinion formation may be different from each other. They satisfy the equation

\begin{equation}
\label{eq.1}
w_l=(1-\Delta w)^{l-1}.
\end{equation}
Therefore, for $\Delta w=0$, all the communities have the same weight of the influence on one's opinion formation. For $\Delta w\ne0$, the maximal weight keeps $w_1=1$ and the average weight decreases with the rise of $L$.

Initially, each individual selects opinion $\sigma_+$ or $\sigma_-$ randomly. In the updating process. For an individual i with opinion $\sigma_+$ (or $\sigma_-$). Firstly, he chooses the $lth$ subnetwork from the total $L$ subnetworks randomly. With probability

\begin{equation}
\label{eq.1}
T_r=\frac{n_{\sigma-}}{n_{\sigma+}+n_{\sigma-}}
\end{equation}
(or $T_r=\frac{n_{\sigma+}}{n_{\sigma-}+n_{\sigma+}}$), in which $n_{\sigma+}$ (or $n_{\sigma-}$) represents the number of individuals with state $\sigma_+$ (or $\sigma_-$) in the community, individual $i$ accepts a dose of the influence of peer pressure, $d_0=w_l$. Or else, individual $i$ does not accept the influence of peer pressure, $d_0=0$. Secondly, individual $i$ calculates the cumulative influence of peer pressure in his memory,

\begin{equation}
\label{eq.1}
D=\Sigma_{m'=0}^{m-1} {(1-\xi)^{m'}}{d_{m'}},
\end{equation}
in which the variables of $d_{m'}$, $m$ and $\xi$ represent a dose of the influence accepted $m'$ times before, the memory length and the time-decaying rate of the influence of peer pressure respectively. For an updating threshold $d_{th}$. If $D>d_{th}$, individual $i$ updates his opinion. Or else, individual i keeps his opinion. The parameter $\xi\in[0,1]$ represents the time-dependent effect of the influence of peer pressure. For $\xi=0$, the accepted influence of peer pressure does not change with time. For $0<\xi<1$, the accepted influence of peer pressure decays with time. Therefore, for given memory length $m$, increasing $\xi$ leads to a decrease in the possible maximal value of the cumulative influence of peer pressure.

The instantaneous magnetization $M$ represents the difference between the frequencies of the individuals with state $\sigma_+$ and state $\sigma_-$, which satisfies the equation

\begin{equation}
\label{eq.1}
M=\frac{\mid N_{\sigma+}-N_{\sigma-}\mid}{N}.
\end{equation}
$M=1$ means that the system has reached the state where all the people have the same opinion (consensus). $M=0$ means that the system has reached the state where half of the people exhibit $\sigma_+$ state and another half of the people exhibit $\sigma_-$ state. We should pay special attention to the conditions for the occurrence of phase transitions between $M=1$ and $M\ne1$ and the relaxation time to reach a consensus in the present work.

\section{Simulation results and discussions}
\label{sec:results}

In the present model, we are especially concerned about the coupled effects of heterogeneity in peer pressure and the decaying rate of the influence on the occurrence of phase transitions between a disordered system and an ordered system. The conditions for the change of magnetization $M$ and convergence time $t_c$ are checked respectively in the following.

\begin{figure}
\includegraphics[width=14cm]{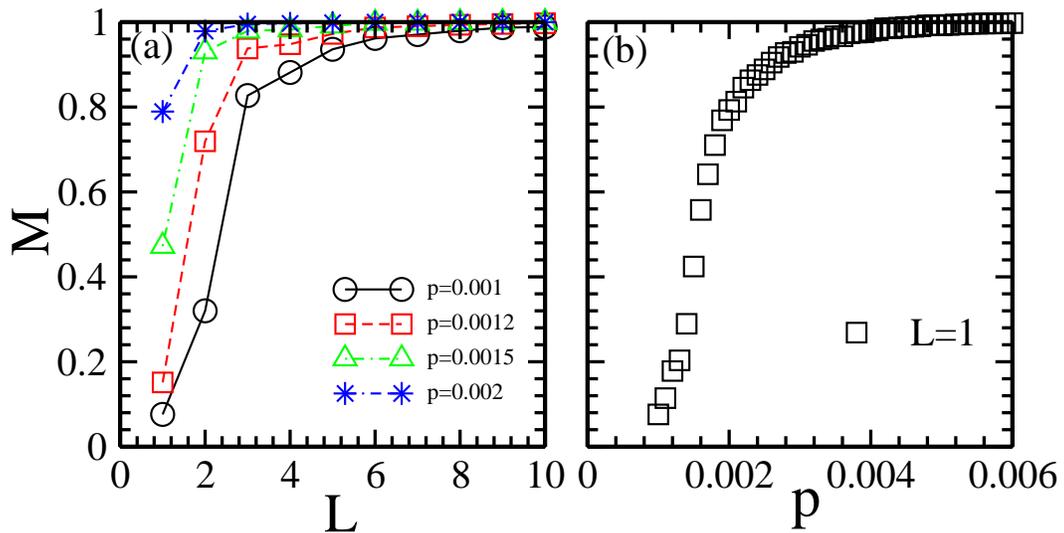}
\caption{\label{fig:epsart}The difference between the number of individuals with opinion $\sigma_+$ and the number of individuals with opinion $\sigma_-$ indicates the degree of coincidence. Shown is the averaged magnetization $M$ (a) as a function of multilayer $L$ for random connection p=0.001 (circles), 0.0012 (squares), 0.0015 (triangles), 0.002 (stars); (b) as a function of random connection $p$ for $L$=1 (squares). Other parameters are: population size $N=1000$, memory length $m=10$, updating threshold $d_{th}=5$, heterogeneity factor $\Delta w=0$, decaying rate $\xi=0$. It can be observed that increasing multilayer $L$ for a fixed $p$ is similar to increasing random connection $p$ for a fixed $L$. The multilayer linkage is beneficial for consensus in homogeneous environment.}
\end{figure}

Figure 1 (a) displays the averaged magnetization $M$ as a function of multilayer
 $L$ for different random connection $p$. For $p=0.001$. As $L$ increases from $L=1$ to $L=3$, $M$ increases quickly from $M\sim0.08$ to $M\sim0.82$. As $L$ increases from $L=3$ to $L=10$, $M$ increases slowly and reaches its maximal value of $M\sim1$.

In fig. 1 (b) we plot the averaged magnetization $M$ as a function of random connection $p$ for $L$=1. As $p$ increases from $p$=0.001 to $p\sim0.004$, $M$ increases sharply with the rise of $p$ and reaches its maximal value of $M\sim1$. A further increase in $p$ has little effect on the change of $M$.

Such results indicate that, as $L$ increases, the individuals in the disconnected component in one subnetwork are more possible to be connected to the others in another subnetwork. In relation to the conditions for consensus, increasing multilayer $L$ for a fixed $p$ is similar to increasing random connection $p$ for a fixed $L$. In homogeneous environment, the multiple linkage is beneficial for consensus.

\begin{figure}
\includegraphics[width=14cm]{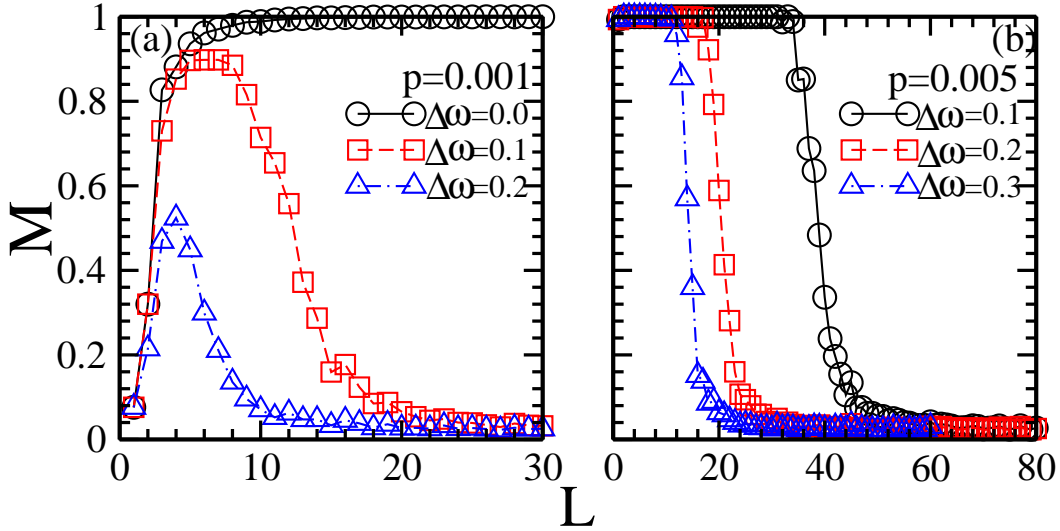}
\caption{\label{fig:epsart} The heterogeneity factor $\Delta w$ indicates the difference in the influence of peer pressure between different layers. Shown is the averaged magnetization $M$ as a function of multilayer $L$ for (a) random connection p=0.001 and heterogeneity factor $\Delta w$=0 (circles), 0.1 (squares), 0.2 (triangles); (b) p=0.005 and $\Delta w$=0.1 (circles), 0.2 (squares), 0.3 (triangles). Other parameters are: population size $N=1000$, memory length $m=10$, dose threshold $d_{th}=5$, decaying rate $\xi=0$. An increase in multilayer $L$ may lead to an increase in the average number of links each individual has and a decrease in the influence of each layer. The former is beneficial for while the latter is detrimental to reaching a consensus.}
\end{figure}

To find out whether the heterogeneity in peer pressure has an impact on opinion formation, in fig. 2 (a) and (b) we plot the averaged magnetization $M$ as a function of multilayer $L$ for different random connection $p$. For $p$=0.001 and $\Delta w$=0.1. As $L$ increases from $L$=1 to $L$=6, $M$ increases from $M\sim$0.08 to $M\sim$0.9. As $L$ increases from $L$=7 to $L$=20, $M$ decreases from $M\sim$0.9 to $M\sim$0.05. A further increase in $L$ has little effect on the change of $M$. Increasing $\Delta w$ leads to an overall decrease in $M$ within the range of $1<L<20$.

For $p$=0.005 and $\Delta w$=0.1. There exists a critical value of $L_c\sim 35$, below which $M$ keeps its maximal value of $M$=1 and above which $M$ decreases with the rise of $L$ and finally reaches its minimal value of $M\sim$0.05. Increasing $\Delta w$ leads to a decrease in $L_c$ but not the changing tendency of $M$ vs $L$.

\begin{figure}
\includegraphics[width=9cm]{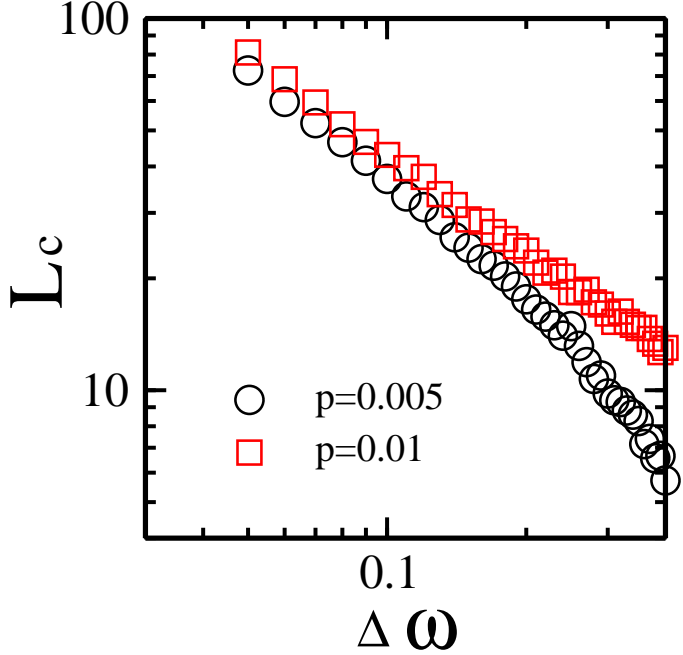}
\caption{\label{fig:epsart} The role of heterogeneity in peer pressure can be reflected in the relationship between the critical point $L_c$ of multilayer and the heterogeneity factor $\Delta w$. Shown is the critical point $L_c$ as a function of heterogeneity factor $\Delta w$ for random connection p=0.005(circles), 0.01(squares). Other parameters are: population size $N=1000$, memory length $m=10$, dose threshold $d_{th}=5$, decaying rate $\xi=0$. It is observed that, for different values of random connection $p$, increasing heterogeneity in peer pressure is detrimental to reaching a consensus. The changing tendency of $L_c$ vs $\Delta w$ is closely related to the average number of connections each individual has.}
\end{figure}

Figure 3 shows the critical point $L_c$ of multilayer as a function of heterogeneity factor $\Delta w$ for different random connection $p$. For $p=0.005$, as $\Delta w$ increases from 0.05 to 0.4, $L_c$ decreases from $L_c\sim70$ to $L_c\sim6$.  Increasing $p$ leads to an overall increase in $L_c$. As we draw a line of best fit on the scattered data, a log-log relation between $L_c$ and $\Delta w$, which satisfies $L_c\sim g(p) (\Delta w)^{-\alpha}$, i.e., $g(p)\sim 2.4$, $\alpha\sim 1.2$ for $p=0.005$ and $g(p)\sim 5.6$, $\alpha\sim 0.89$ for $p=0.01$, is found.

The results in fig.2 and fig.3 indicate that the existence of multi-communities is not always beneficial for people to reach an agreement. In inhomogeneous environment, the multilayer is detrimental to consensus.

\begin{figure}
\includegraphics[width=14cm]{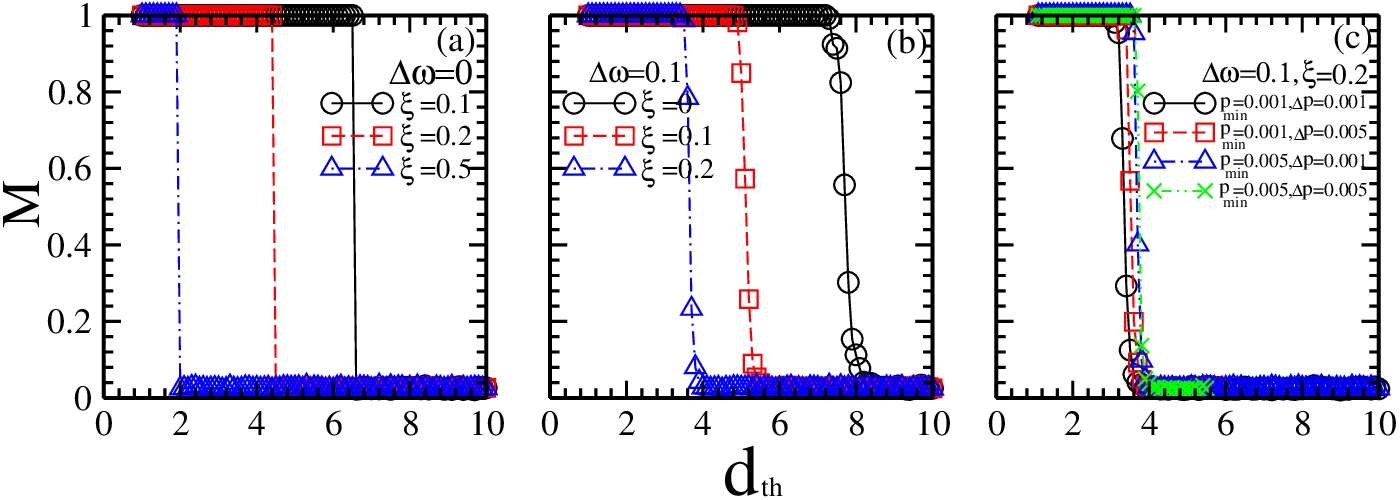}
\caption{\label{fig:epsart}The decaying rate $\xi$ indicates the difference in the influence of peer pressure accepted at different time. The updating threshold $d_{th}$ denotes the conditions for the possible occurrence of state updating. Shown is the averaged magnetization $M$ as a function of updating threshold $d_{th}$ for (a) heterogeneity factor $\Delta w=0$, random connection $p=0.005$ and decaying rate $\xi$= 0.1 (circles), 0.2 (squares), 0.5 (triangles); (b) $\Delta w=0.1$, $p=0.005$ and $\xi$= 0 (circles), 0.1 (squares), 0.2 (triangles). (c) $\Delta w=0.1$, $\xi=0.2$, minimal connection probability $p_{min}=0.001$, minimal probability difference $\Delta p=0.001$ (circles); $p_{min}=0.001$, $\Delta p=0.005$ (squares); $p_{min}=0.005$, $\Delta p=0.001$ (triangles); $p_{min}=0.005$, $\Delta p=0.005$ (times). Other parameters are: population size $N=1000$, memory length $m=10$, multilayer $L=10$.  It is observed that, compared with the situation where there is no attenuation, the time-decaying effect leads to a decrease in the maximal value of the updating threshold, above which an update is impossible to occur. An increase in the average number of connections each individual owns leads to an increase in the critical point $d_{th}^c$.}
\end{figure}

For each individual, he accepts the influence of different communities at different time. In order to find out the time-decaying effect of the influence of peer pressure on consensus, in fig. 4 (a) and (b) we plot the averaged magnetization $M$ as a function of updating threshold $d_{th}$ for different decaying rate $\xi$. For $\Delta w=0$. There exists a critical point $d_{th}^c$ of updating threshold, below which $M$ keeps its maximal value $M$=1 and above which $M$ keeps its minimal value $M\sim0.05$. An increase in $\xi$ only leads to a decrease in $d_{th}^c$ but not the maximal and minimal values of $M$. Similar results are also found for $\Delta w=0.1$.

Figure 4 (c) shows the averaged magnetization $M$ as a function of updating threshold $d_{th}$ for a multilayer network with different subnetwork topology. For the case that different subnetworks have different random connection $p$, the changing tendency of $M$ vs $d_{th}$ is similar to the case that different subnetworks have the same random connection $p$. An increase in the average number of connections leads to an increase in the critical point $d_{th}^c$.

\begin{figure}
\includegraphics[width=9cm]{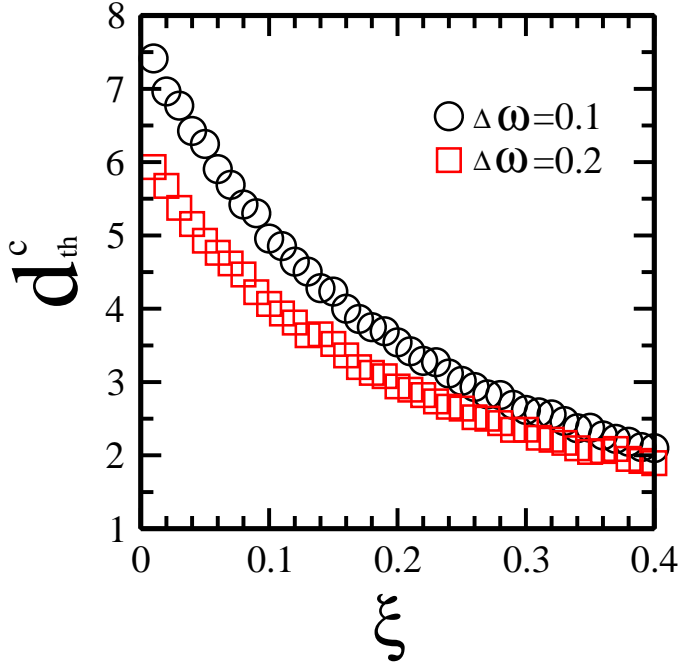}
\caption{\label{fig:epsart} The role of attenuation of the influence of peer pressure can be reflected in the relationship between the critical point $d_{th}^c$ of updating threshold and the decaying rate $\xi$. Shown is the critical point $d_{th}^c$ as a function of decaying rate $\xi$ for heterogeneity factor $\Delta w$=0.1(circles), 0.2(squares). Other parameters are: population size $N=1000$, memory length $m=10$, multilayer $L=10$, random connection $p=0.005$. It is observed that, for different values of heterogeneity factor $\Delta w$, increasing decaying rate $\xi$ is detrimental to reaching a consensus. The changing tendency of $d_{th}^c$ vs $\xi$ is closely related to the heterogeneity factor $\Delta w$.}
\end{figure}

Figure 5 shows the critical point $d_{th}^c$ of updating threshold as a function of decaying rate $\xi$ for different values of heterogeneity factor $\Delta w$. For $\Delta w$=0.1. As $\xi$ increases from $\xi$=0.01 to $\xi$=0.4, $d_{th}^c$ decreases from $d_{th}^c\sim 7.3$ to $d_{th}^c\sim 2.1$. Increasing $\Delta w$ leads to an overall decrease in $d_{th}^c$. As we draw a line of best fit on the scattered data, a reciprocal relation between $d_{th}^c$ and $\xi$, which satisfies $d_{th}^c\sim\frac{1}{a+b\xi}$, i.e., $a\sim 0.12$, $b\sim 0.80$ for $\Delta w$=0.1 and $a\sim 0.15$, $b\sim 0.93$ for $\Delta w$=0.2, is found.

The results in fig.4 and fig.5 indicate that, the time-decaying mechanism deepens the influence of heterogeneity in peer pressure and is detrimental to consensus.

\begin{figure}
\includegraphics[width=14cm]{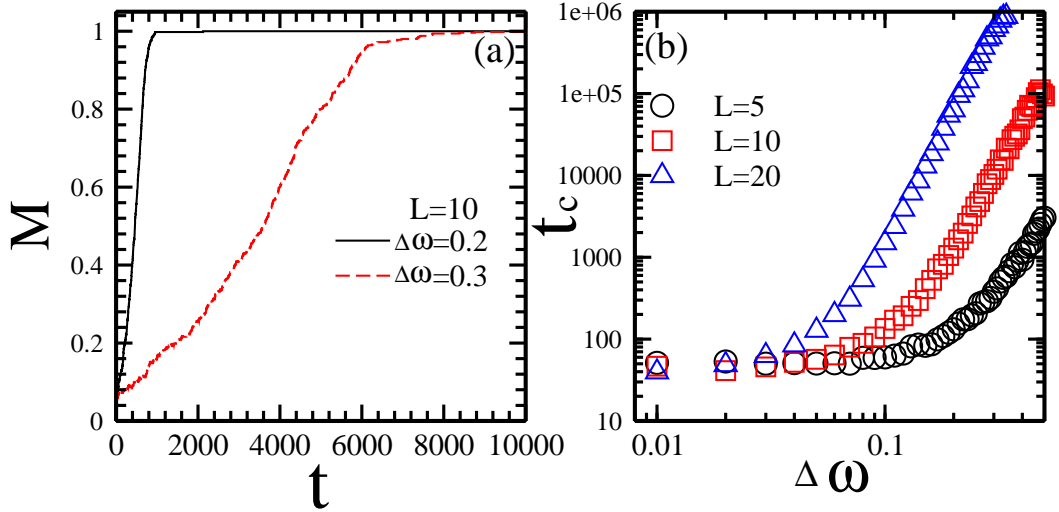}
\caption{\label{fig:epsart}The convergence time indicates the degree of difficulty in reaching a consensus. (a) The time-dependent value of magnetization $M$ for multilayer $L=10$ and heterogeneity factor $\Delta w$=0.2(line), 0.3(slash). (b) The convergence time $t_c$ as a function of heterogeneity factor $\Delta w$ for multilayer $L$=5 (circles), 10 (squares), 20 (triangles). Other parameters are: population size $N=1000$, memory length $m=10$, random connection $p=0.005$, decaying rate $\xi$=0, updating threshold $d_{th}$=5. It is observed that both increasing multilayer $L$ and heterogeneity factor $\Delta w$ lead to a rise of the convergence time.}
\end{figure}

The above results indicate that the evolutionary dynamics in the present model is closely related to the heterogeneity in peer pressure and the decaying rate of the influence. In order to get a deep understanding of the role of parameters $\Delta w$ and $\xi$ in the occurrence of consensus, in the following, the relationship between the convergence time and the variables of $\Delta w$ and $\xi$ is investigated.

Figure 6 (a) shows the time-dependent behavior of magnetization $M$ for different heterogeneity factor $\Delta w$. For $\Delta w$=0.2, magnetization $M$ rises sharply from $M\sim0.05$ to $M\sim1$ within the range of $0<t<1000$. The convergence time $t_c\sim1000$ is found. For a larger $\Delta w$, the changing tendency of $M$ vs $t$ becomes ease and the convergence time increases significantly. From fig.6(b) we find that, for a large value of $\Delta w$, increasing $\Delta w$ and $L$ both lead to a rise of $t_c$.

Comparing the results in fig. 2 and fig. 3 with the results in fig. 6, we find that the occurrence of phase transitions of $M$ vs $L$ should result from the rise of the relaxation time to the final steady state.

\begin{figure}
\includegraphics[width=14cm]{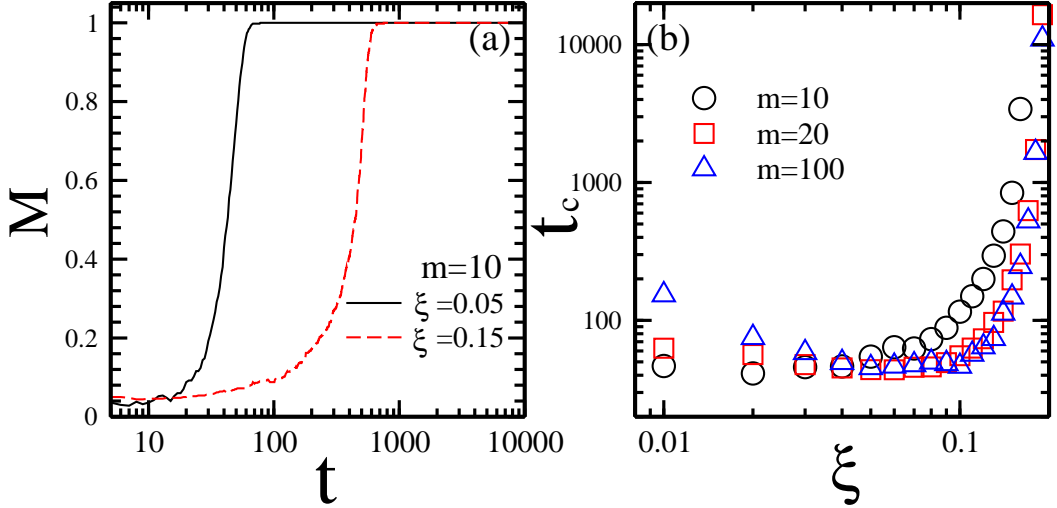}
\caption{\label{fig:epsart}As we consider the role of attenuation in the convergence time, memory length may have a great impact on the attenuation pattern. Shown is (a) the time-dependent value of magnetization $M$ for decaying rate $\xi$=0.05(line), 0.15(slash); (b) the convergence time $t_c$ as a function of decaying rate $\xi$ for memory length $m$=10 (circles), 20 (squares), 100 (triangles). Other parameters are: population size $N=1000$, multilayer $L=10$, random connection $p=0.005$, heterogeneity factor $\Delta w$=0, dose threshold $d_{th}$=5. It is observed that the changing tendency of $t_c$ vs $\xi$ is closely related to the memory length $m$. For a quite large value of memory length, a critical decaying rate, at which it needs much lower relaxation time to reach a consensus, is found.}
\end{figure}

Figure 7 (a) shows the time-dependent behavior of magnetization $M$ for different decaying rate $\xi$. For $\xi$=0.05, the magnetization $M$ fluctuates around $M\sim 0.05$ within the range of $0<t<20$. $M$ rises sharply from $M\sim0.05$ to $M\sim1$ within the range of $20<t<70$. The convergence time $t_c\sim70$ is found. For $\xi$=0.15, the magnetization $M$ increases slowly from $M\sim 0.05$ to $M\sim 0.1$ within the range of $0<t<100$. $M$ rises sharply from $M\sim0.1$ to $M\sim1$ within the range of $100<t<600$. The convergence time $t_c\sim600$ is found.

Figure 7 (b) shows the convergence time $t_c$ as a function of the decaying rate $\xi$ for different memory length $m$. It is observed that, for a small value of memory length $m$, an increase in $\xi$ firstly has little impact on the change of $t_c$ and then leads to a sharp rise of $t_c$. For a large value of memory length $m$, an increase in $\xi$ firstly leads to a decrease in $t_c$ and then a sharp rise of $t_c$. For $m=10$, as $\xi$ increases from 0.01 to 0.08, $t_c$ increases from $t_c\sim$45 to $t_c\sim$100. As $\xi$ increases from 0.08 to 0.2, $t_c$ increases sharply from $t_c\sim$100 to infinity. An increase in $m$ leads to an overall increase in $t_c$ within the range of $0.01<\xi<0.04$ and an overall decrease in $t_c$ within the range of $0.05<\xi<0.2$. For $m=100$, as $\xi$ increases from 0.01 to 0.1, $t_c$ decreases from $t_c\sim$160 and reaches its minimal value of $t_c\sim$45. As $\xi$ increases from 0.1 to 0.2, $t_c$ increases sharply from $t_c\sim$45 to infinity.

The results in fig. 6 and fig. 7 indicate that the occurrence of nonconsensus state in the present model should be related to the rise of the relaxation time to the final steady state. For large values of $\Delta w$ and $\xi$, the relaxation time to the final steady state becomes infinity, which makes it impossible for the system to reach a consensus.

\begin{figure}
\includegraphics[width=14cm]{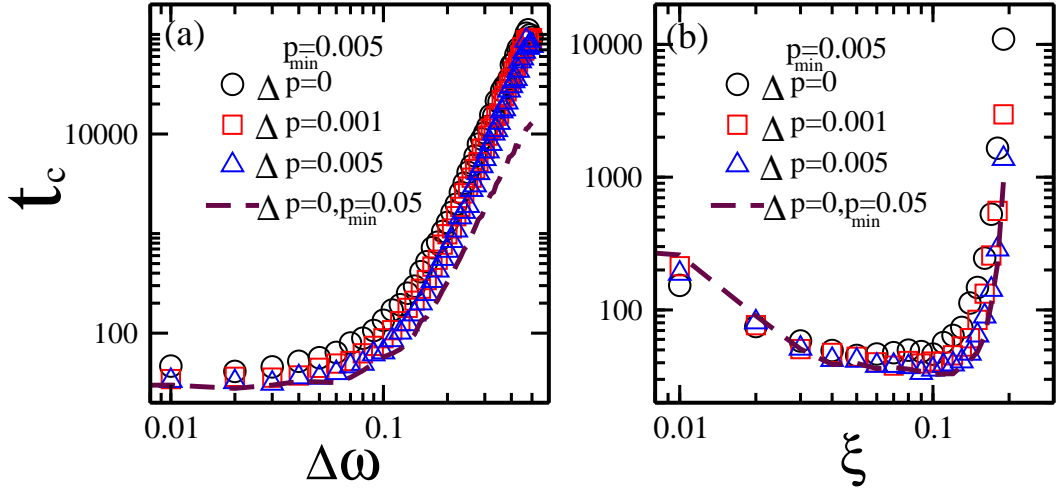}
\caption{\label{fig:epsart}Different subnetworks in a multilayer network may have different random connection $p$. Shown is the convergence time $t_c$ (a) as a function of heterogeneity factor $\Delta w$ for memory length $m$=10, decaying rate $\xi=0$, minimal connection probability $p_{min}=0.005$, minimal probability difference $\Delta p=0$ (circles), 0.001(squares), 0.005 (triangles); (b)as a function of decaying rate $\xi$ for $m$=100, $\Delta w$=0, $p_{min}=0.005$, $\Delta p=0$ (circles), 0.001 (squares), 0.005 (triangles). Other parameters are: population size $N=1000$, multilayer $L=10$,  updating threshold $d_{th}$=5. The slash line is plotted for comparison. It is observed that the changing tendency of $t_c$ vs $\Delta w$ or $t_c$ vs $\xi$ is related to the average number of connections. Within the whole range of $0.01<\Delta w<0.5$, an increase in the average number of connections leads to an overall decrease in the convergence time. Within the range of $0.01<\xi<0.03$, an increase in the average number of connections leads to an increase in the convergence time. Within the range of $0.03<\xi<0.2$, an increase in the average number of connections leads to a decrease in the convergence time. }
\end{figure}

Figure 8 (a) shows the convergence time $t_c$ as a function of heterogeneity factor $\Delta w$ for a multilayer network in which the subnetworks have different random connection $p$. It is observed that the existence of the difference in subnetwork topology has little impact on the convergence time. An increase in the average number of connections each individual owns leads to a decrease in $t_c$ within the whole range of $0.01<\Delta w<0.5 $, which is similar to the results found in fig. 6 (b).

Figure 8 (b) shows $t_c$ as a function of $\xi$ for a multilayer network in which the subnetworks have different random connection $p$. It is also observed that the existence of the difference in subnetwork topology has little impact on the convergence time. Within the range of $0.01<\xi<0.03$, an increase in the average number of connections leads to an increase in the convergence time. Within the range of $0.03<\xi<0.2$, an increase in the average number of connections leads to a decrease in the convergence time, which is similar to the results found in fig. 7 (b).

The results in fig. 4 (c) and fig. 8 indicate that the difference in subnetwork topology has little impact on opinion dynamics. An increase in the average number of connections have an impact on the system behavior.

\section{Theoretical analysis}
\label{sec:analysis}
\subsection{\label{subsec:levelA}relationship between the critical point of updating threshold and the decaying rate of the influence of peer pressure}

For an individual $i$ with opinion $\sigma_+$, the probability of accepting the influence of peer pressure is proportional to the frequency of individuals with state $\sigma_-$ in the community, which satisfies the equation $T_r^i=\frac{n_{\sigma-}}{n_{\sigma+}+n_{\sigma-}}$. For a given system with memory length $m$ and multilayer $L$, the cumulative influence of peer pressure is $D_i=\Sigma_{m'=1}^m (1-\xi)^{m'-1}d_{m'}$, in which $d_{m'}=(1-\Delta w)^{l-1}d_0$, $d_0=1$ with probability $T_r^i$ and $d_0=0$ with probability $1-T_r^i$. On condition that $D_i\ge d_{th}$, individual $i$ switches his opinion $\sigma_+$ to opinion $\sigma_-$.

Consider an initial condition where each individual's state is randomly allocated, $\sigma_+$ or $\sigma_-$. For the system with average community size $\bar n$ and average weight $\bar w$ of the influence of peer pressure. On condition that there is no time-decaying effect. The average value of the influence of peer pressure is

\begin{equation}
\label{eq.1}
\bar{D_i}=m\bar{T_r^i}\bar{w_l}.
\end{equation}
The maximal value of the influence of peer pressure is

\begin{equation}
\label{eq.1}
D_i^{max}=mT_r^{max}.
\end{equation}
If the inequality ${D^{max}_i}< d_{th}$ is always satisfied, the frequency of individuals with state $\sigma_+$ will not change with time. Therefore, the critical point of the updating threshold is obtained,

\begin{equation}
\label{eq.1}
d^c_{th}=mT_r^{max}.
\end{equation}
On condition that the time-decaying effect is considered. For an average value of the time-decaying parameter $\bar x$. The average value of the influence of peer pressure becomes

\begin{equation}
\label{eq.1}
\bar{D_i}=m\bar x\bar{T_r^i}\bar{w_l}.
\end{equation}
The maximal value of the influence of peer pressure becomes

\begin{equation}
\label{eq.1}
{D^{max}_i}=(1-\xi)^0+(1-\xi)^1+...+(1-\xi)^{mT_r^{max}-1}.
\end{equation}
Therefore, the consensus threshold becomes

\begin{equation}
\label{eq.1}
d^c_{th}=\frac{[1-(1-\xi)^{mT_r^{max}}]}{\xi}.
\end{equation}

From the above analysis we find that, for the initial condition $n_{\sigma +}=n_{\sigma -}=\frac{1}{2}n$, the state updating probability is determined by the difference between $\bar{D_i}$ and $d_{th}$. Before a consensus is reached, the state updating probability is determined by the difference between $D^{max}_i$ and $d_{th}$. The critical point $d_{th}^c$ decreases with the rise of $\xi$, which is in accordance with the simulation data in fig. 5.

\subsection{\label{subsec:levelB} Relationship between the convergence time and the decaying rate of the influence of peer pressure}

In the present model, the convergence time $t_c$ is related to the time $t_{meta}$ to escape from a metastable state and the time $t_{con}$ to reach a consensus, $t_c=t_{meta}+t_{con}$. $t_{con}$ is mainly determined by the difference in the frequencies between two opposed opinions. As the number of individuals with opinion $\sigma+$ is equal to the number of individuals with opinion $\sigma-$, i.e., $n_{\sigma +}=n_{\sigma -}=\frac{1}{2}n$, the convergence time is mainly determined by $t_{meta}$.

The time $t_{meta}$ is related to the following two factors: the time interval $\Delta T$ within which an update occurs and the time interval $\Delta t$ within which two opposed opinions update sequentially. For a given updating threshold $d_{th}$, $\Delta T$ is determined by the cumulative influence of peer pressure $D$, which is determined by memory length $m$ and time-decaying rate $\xi$. Suppose the probability of accepting the influence of peer pressures is $P_{accept}$ and the accepted influence is $d_0=1$. On condition that $n_{\sigma +}=n_{\sigma -}=\frac{1}{2}n$, the updating probability is satisfied with the equation

\begin{equation}
\label{eq.1}
P_{update}=(\frac{1}{2})^{d_{th}}+\Sigma_{m'=d_{th}+1}^m C_{m'-1}^{d_{th}-1}(\frac{1}{2})^{m'}.
\end{equation}
A large updating probability means a small $\Delta T$. Suppose

\begin{equation}
\label{eq.1}
\Delta T\sim\frac{1}{(\frac{1}{2})^{d_{th}}+\Sigma_{m'=d_{th}+1}^m C_{m'-1}^{d_{th}-1}(\frac{1}{2})^{m'}},
\end{equation}
we find that $\Delta T$ decreases with the rise of memory length $m$.

$\Delta t$ is determined by the time interval between the updates of two opposed opinions. Suppose that the difference between the updating probabilities of $P_{\sigma_+\to\sigma_-}$ and $P_{\sigma_-\to\sigma_+}$ is quite small and $\Delta t$ satisfies the relation

\begin{equation}
\label{eq.1}
\Delta t\sim C_{m-1}^{d_{th}-1}(\frac{1}{2})^{m}.
\end{equation}
We find that, for a given $d_{th}$, $\Delta t$ firstly increases and then decreases with the rise of $m$.

The time $t_{meta}$ in the metastable state increases with the rise of $\Delta T$ and decreases with the rise of $\Delta t$. Suppose that they satisfy the relation

\begin{equation}
\label{eq.1}
t_{meta}\sim\frac{\Delta T}{\Delta t},
\end{equation}
we get

\begin{equation}
\label{eq.1}
t_{meta}\sim\frac{1}{[(\frac{1}{2})^{d_{th}}+\Sigma_{m'=d_{th}+1}^m C_{m'-1}^{d_{th}-1}(\frac{1}{2})^{m'}][C_{m-1}^{d_{th}-1}(\frac{1}{2})^{m}]}.
\end{equation}
From the above analysis we find that, for a given $d_{th}$, within the range where $m$ is not quite large, increasing $m$ leads to a rise of $(\frac{1}{2})^{d_{th}}+\Sigma_{m'=d_{th}+1}^m C_{m'-1}^{d_{th}-1}(\frac{1}{2})^{m'}$ and a rise of $C_{m-1}^{d_{th}-1}(\frac{1}{2})^{m}$, the coupling of which results in a decrease in $t_{meta}$. Within the range where $m$ is large, increasing $m$ leads to a rise of $(\frac{1}{2})^{d_{th}}+\Sigma_{m'=d_{th}+1}^m C_{m'-1}^{d_{th}-1}(\frac{1}{2})^{m'}$ and a decrease in $C_{m-1}^{d_{th}-1}(\frac{1}{2})^{m}$, the coupling of which results in an increase in $t_{meta}$.

Consider the role of $\xi$ in the change of $t_{meta}$. Increasing $\xi$ leads to a decrease in the possible maximal value of $D$, which is similar to the results obtained by reducing $m$. Therefore, for a quite large $m$ and a small $\xi$, increasing $\xi$ is similar to reducing $m$ within the range where $m$ is large. Increasing $\xi$ leads to a decrease in $t_{meta}$. For a large $\xi$, increasing $\xi$ is similar to reducing $m$ within the range where $m$ is not quite large. Increasing $\xi$ leads to an increase in $t_{meta}$.

The changing tendencies of $t_{meta}$ vs $m$ and $t_{meta}$ vs $\xi$ in theoretical analysis are in accordance with the simulation results of $t_c$ vs $m$ and $t_c$ vs $\xi$ in fig. 7.

\section{Summary}
\label{sec:summary}
By incorporating multi-community structures and time-decaying memory into the original voter model, we propose a generalized voter model to study the cumulative effect of the influence of peer pressure on the evolution of opinions. The inhomogeneity in peer pressure has a great impact on the occurrence of transition between an ordered system and a disordered system. The transition point is determined by the coupling of the heterogeneity in peer pressure and the decaying rate of the influence. An increase in heterogeneity or decaying rate is detrimental to consensus. There exists a critical point of the decaying rate, below which the convergence time decreases with the rise of the decaying rate while above which the convergence time increases with the rise of the decaying rate. A mean-field analysis indicates that the occurrence of phase transitions in the present model should be governed by the relaxation time to the final steady state.

F. Battiston et al. have also studied the heterogeneity effect depending upon the voter model\cite{battiston1}. They have found that the relative weight assigned to social pressure, internal coherence and external mean-field force has a great impact on the occurrence of phase transitions between a disordered system and an ordered system, which is similar to the role of the heterogeneous influence of peer pressure in our model. The difference between the two models is that neither the cumulative effect nor the time-decaying effect has been considered in their model. H. U. Stark et al. have also studied the time-decaying influence of peer pressure in the voter model\cite{stark1,stark2}. They have found that, for an intermediate inertia rate, the convergence time can reach its minimal value, which is similar to the role of memory length in our model. The difference between the two models is that the heterogeneous effect and the cumulative effect have not been considered in their model. Therefore, compared with the original voter model, the model proposed in the present work can help us explore the comprehensive effect of multi-factors on opinion formation.

In the future, the coupled effects of a multi-layer network and time-decaying memory will be extensively studied in a variety of diffusion systems. The coupling of different kinds of subnetworks, such as dynamic network and scalefree network,
will be considered in depth.

\section*{Acknowledgments}
This work is the research fruits of National Natural Science Foundation of China (Grant Nos. 71371165, 61503109, 11175079, 71471161,71171176), Research Project of Zhejiang Social Sciences Association (Grant No. 2014N079), Research Project of Generalized Virtual Economy(Grant No. GX2015 -1004(M)), Jiangxi Provincial Young Scientist Training Project (Grant No. 2013 3BCB23017).





\bibliographystyle{model1-num-names}



\end{document}